\documentclass[fleqn,twoside]{article}
\usepackage{espcrc2}

\usepackage{graphicx}

\newcommand{\AmS}{{\protect\the\textfont2
  A\kern-.1667em\lower.5ex\hbox{M}\kern-.125emS}}

\newcommand{\hmu}{\hat{\mu}}
\newcommand{\hnu}{\hat{\nu}}
\newcommand{\hrho}{\hat{\rho}}

\def\tr{{\rm ~tr}\,}

\title{Super Yang-Mills Theory on Lattice and the Transformation}

\author{K. Itoh\address{Faculty of Education, 
        Niigata University, Niigata 950-2181, Japan},
        M. Kato\address{Institue of Physics, University of Tokyo, Komaba, 
                        Meguroku, Tokyo 153, Japan},
        H. Sawanaka\address[DPNU]{Department of Physics, Niigata University, 
        Niigata 950-2181, Japan},
        H. So\addressmark[DPNU]\thanks{Talk presented by H. So. This work was 
        supportted in part by Grants-in-Aid for Scientific Research No. 12640259
          from the Japan Society for the Promotion of Science.}
        and
        N. Ukita\addressmark[DPNU]}
       
\begin{document}

\begin{abstract}
We present a new lattice super Yang-Mills theory and its SUSY
transformation. After our formulation of the model in a fundamental
lattice, it is extended to the whole lattice with a substructure of modulo 2.
\vspace{1pc}
\end{abstract}

\maketitle

\section{INTRODUCTION}

A lattice formulation is one of the most powerful tools for a
nonperturbative study. We present here a completely new formulation of
the super Yang-Mills theory on lattice with an exact fermionic symmetry.

There are several pionieering attempts for contructing super Yang-Mills
theories on lattice.  The first approach, the Wilson fermion and mass
fine-tuning procedure, is a well-simulated method \cite{CV,MV}.  It is
based on an observation that, in the perturbative expansion, a gaugino
mass is the unique relevant operator which breaks supersymmetry around
the SUSY fixed point \cite{CV,TG}.  
Nevertheless, nobody knows whether this observation is nonperturbatively
correct and there exists the SUSY fixed point in the Wilson fermion theory. 

Second is an appoach with the staggered fermion \cite{AGZ,AZ}. 
This formulation may allow us to have a correspondence
between fermion and boson.  However, the counting of the degrees of
freedom is still difficult for Yang-Mills theory and the full theory has
not been constructed until now.

Third approach is based on the domain wall fermion \cite{KS}.  This
approach naturally has a vanishing mechanism of the gaugino mass,
although, in order to impose a Majorana condition, one must introduce a
nonlocal condition along the fifth direction.  A simulation has also
been done \cite{FKV}.

Our new formulation may look like the second approach in some aspects.
However, the crucial difference is the presence of an exact fermionic
symmetry in our formulation.  No fine-tuning is required for the
symmetry and we show that it is not BRS-like.  

We start our consideration with a model defined on a fundamental
lattice.  The model will be called as the one-cell model.  Then we show
that the model may be extended to the whole space while keeping the
fermionic symmetry found for the one-cell model.  The extended model has
a slightly unusual lattice structure, which will be explained later.

\section{LATTICE SUSY TRANSFORMATION FOR GAUGE AND FERMION}

In this section, we present the lattice action and SUSY transformations
for the one-cell model.  Each coordinate of a site takes only 0 or 1 in
the unit of $a$ (lattice constant).

Our action of gauge and fermion system is a normal one except the facts
that it is resricted on a fundamental lattice and fermi fields, $\psi$,
are real.

$$
S_g = -\beta \sum_{n,\mu\nu}\tr (U_{n(\mu\nu),\mu\nu} + U_{n(\mu\nu),\nu\mu}) 
$$
\begin{eqnarray*}
S_f = &\sum\limits_{n,\rho}b_{\rho}(n(\rho))\tr
(\psi_{n(\rho)} U_{n(\rho),\rho}\psi_{n(\rho)+\hrho}  U_{n(\rho),\rho}^{\dagger}\\
 & - \psi_{n(\rho)+\hrho} U_{n(\rho),\rho}^{\dagger}\psi_{n(\rho)} U_{n(\rho),\rho})
\end{eqnarray*}
\noindent where $n(\mu\nu)$ is a base point for the plaquette
$(n,\mu\nu)$ and a symbol $n(\rho)$ implies $n_{\rho} =0$ for the $\rho$
coordinate.  The lattice has one plaquette in D=2 case and six
plaquettes in D=3. The coefficients $b_{\rho}$ of the fermion action
appear similarly to the usual staggered fermion.  However the fermion
here is the real fermion.

We also use the symbol $\rho(n)$ to indicate either ${\hat \rho}$ or
$-{\hat \rho}$, allowed inside the cell when we start from the site $n$. 
The SUSY transformation of gauge fields is defined as
\begin{eqnarray*}
\delta U_{n(\mu),\mu} = &\sum_{\rho}(\alpha_{n(\mu),\mu}^{\rho(n)} 
\xi_{n(\mu)}^{\rho(n)}   U_{n(\mu) ,\mu} \\
 &+ ~U_{n(\mu),\mu} 
\alpha_{n(\mu)+\hmu,\mu}^{\rho(n)} 
\xi_{n(\mu)+\hmu}^{\rho(n)})
\end{eqnarray*}
\noindent where $\xi_n^{\mu}$ is a gauge-covariantly translated fermion,
$U_{n,\mu}\psi_{n+\hmu} U^{\dagger}_{n,\mu}$ and $\alpha$ is a
grassmannian parameter.

For fermi fields, we write the transformation as
$$
\delta \psi_{n} = \sum_{0 < \mu < \nu}C_{n}^{(\mu\nu)(n)}(U_{n,(\mu\nu)(n)}  
- U_{n,(\nu\mu)(n)}) 
$$
\noindent
where $(\mu\nu)(n)$  implies  $(-)^{n_{\mu}} \hmu (-)^{n_{\nu}}\hnu$  
 and $C_n^{\mu\nu}$ is  a grassmannian parameter.

\section{SUSY INVARIANCE OF OUR THEORY}

Now, we derive relations between the introduced parameters, which are
required for the SUSY invariance.  They come from four important
conditions.

\vspace{0.3 cm}
\noindent
\fbox{Condition (1) $\delta_{U} S_f = 0 $}
\vspace{0.2 cm}

This is the invariance of the fermion action under gauge field variation. 
It is nothing but the vanishing condition of fermion cubed terms:
$$
b_{\rho}(n) \alpha_{n,\rho}^{\mu} +  b_{\mu}(n)\alpha_{n,\mu}^{\rho} =0.
$$
\noindent
Note that this condition is consistent with the staggered Majorana fermion;
$$
b_{\mu}(n) = (-)^{|n|}\eta_{\mu}(n) b.
$$
\noindent
Here $b$ is a constant and the constraint on $\alpha$ is
$$
\alpha_{n,\nu}^{\mu} = - \frac{b_{\mu}(n)}{b_{\nu}(n)}\alpha_{n,\mu}^{\nu}. 
$$
\noindent
In particular, we find that any diagonal element vanishes
$$
\alpha_{n,\rho}^{\rho} = 0. 
$$

\vspace{0.3 cm}
\noindent
\fbox{Condition (2) $\delta S = \delta_U S_g + \delta_{\psi} S_f = 0  $}
\vspace{0.2 cm}

By satisfying the first and second conditions altogether, the total
action becomes invariant under the SUSY transformation.  The second
condition produces two types of graphs, i.e. those with three different
indices and those with two different indices.
 
\begin{figure}[htbp]
\includegraphics[width=3 cm,height=3 cm]{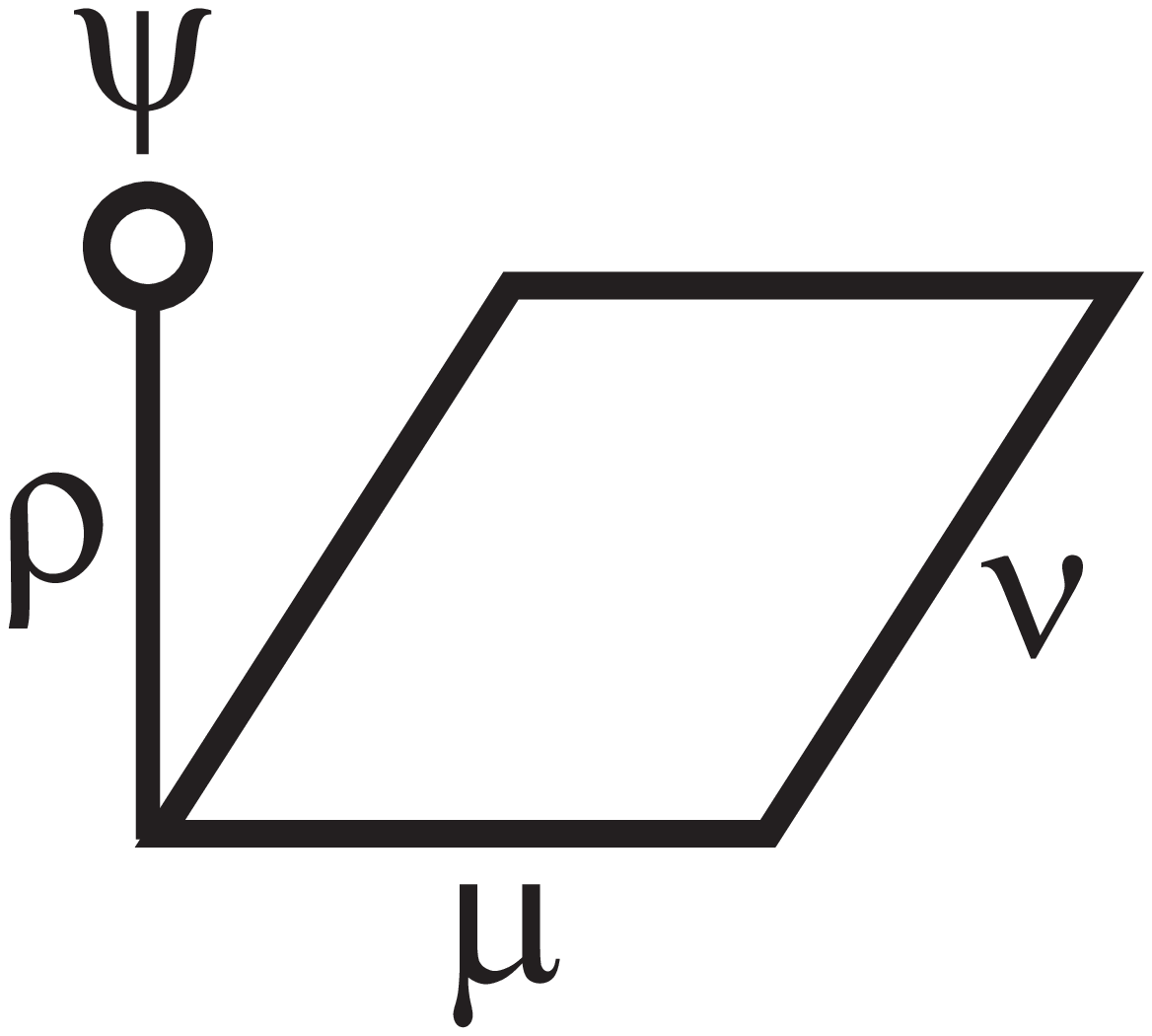}
\includegraphics[width=3 cm,height=2.1 cm]{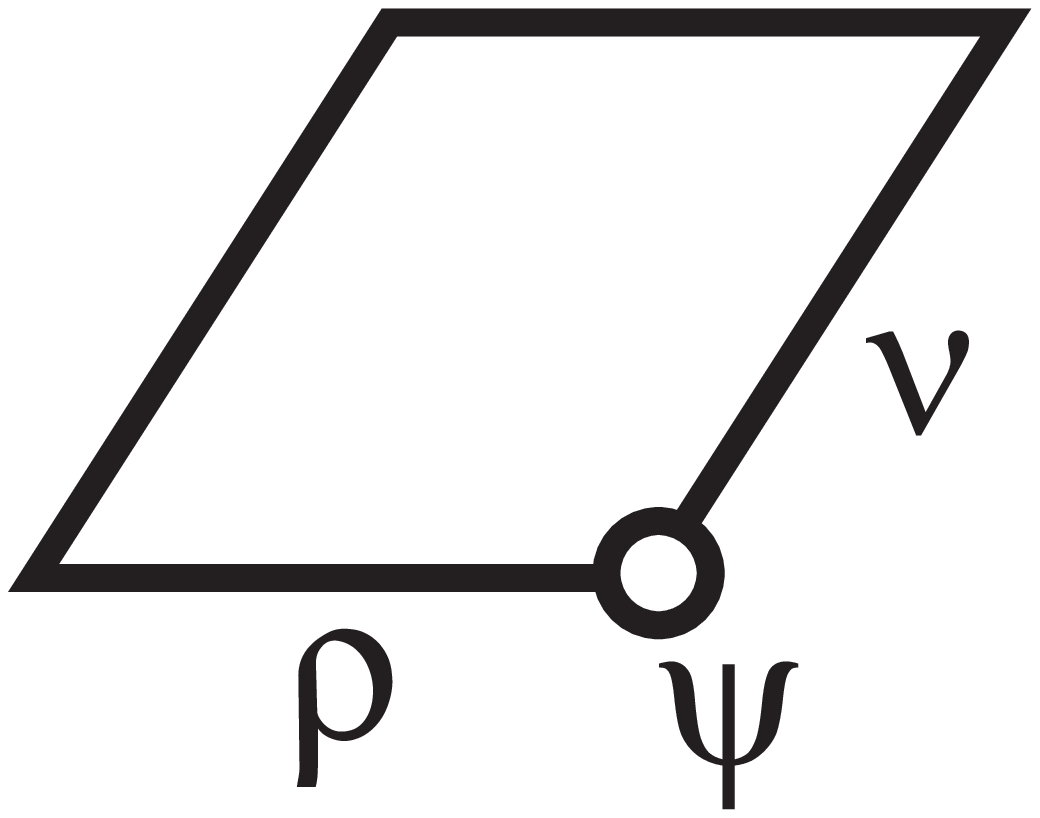}
\caption{Two kinds of graphs}
\end{figure}

\vspace{-0.5 cm}
Corresponding to two types of graphs, we find the relations,

$$
b_{\rho}(n) C_n^{(\mu\nu)(n)} = \beta [(-)^{n_{\mu}}\alpha_{n,\mu}^{\rho}- 
(-)^{n_{\nu}}\alpha_{n,\nu}^{\rho}],  
$$
\noindent
$$
b_{\rho}(n)C_{n+\hrho}^{(\nu-\rho)} + b_{\nu}(n)C_{n+\hnu}^{(-\nu\rho)} = 
-\beta (\alpha_{n+\hrho,\nu}^{-\rho} - \alpha_{n+\hnu,\rho}^{-\nu}).
$$

\vspace{0.3 cm}
\noindent
\fbox{Condition (3) $\delta_U S_g \ne 0 $}
\vspace{ 0.2 cm}

Third one is to confirm that this symmetry is not BRS-like.  We see that
this condition is trivially satisfied because the left hand sides of the
relations obtained from the second condition are nonzero.

\vspace{0.3 cm}
\noindent
\fbox{Condition (4) Invariance of Path Integral Measure}
\vspace{0.2 cm}

The condition is not so trivial.  We consider in two steps.

(A)~ In the first step, only fermion variables change.

$$
U^{'}_{n,\mu} = U_{n,\mu} 
$$
$$
\psi^{'}_n = \psi_n + C_n^{\mu\nu}(U_{n,\mu\nu} - U_{n,\nu\mu})
$$

(B)~ In the second step only gauge field variables change.

$$ U^{''}_{n,\mu} = e^{\alpha_{n,\mu}\cdot \xi^{'}_n}U^{'}_{n,\mu}
e^{\alpha_{n+\hmu,\mu}\cdot \xi^{'}_{n+\hmu}} $$ $$ \psi^{''}_n=
\psi^{'}_n $$ where $\xi_n^{'\rho}= U^{'}_{n,\rho}\{ \psi^{'}_{n+\hrho} -
C_{n+\hrho}^{\mu\nu}(U^{'}_{n+\hrho,\mu\nu} - U^{'}_{n+\hrho,\nu\mu} )
\} U_{n,\rho}^{'\dagger} $.  It is easy to show that Jacobian $J_{\rm A} =
1$.  But $J_{\rm B} =1$ implies $\alpha_{n,\mu}^{\mu} =0$.  This is
consistent with the solution for the first condition.
We have completed the check of the SUSY invariance of our one-cell model.
\section{MULTI-CELL MODEL}

Now, we extend one-cell model to the whole space.  If we do it naiively,
we encounter $O(a)$ reduced SUSY not $O(1)$.  So we have to consider a
new type lattice theory.  Let us decompose the whole lattice into three
units.

(B)~In a blue unit,  a one-cell model lives.

(R)~In a red unit,  another one-cell model lives.

(Bl)~ In blank, there are no cell models.

Our total action is obtained by summing over blue and red cells.


\begin{figure}[htbp]
\includegraphics[scale=0.4]{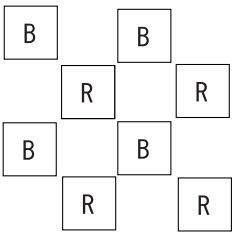}
\caption{A multi-cell pattern}
\end{figure}

\vspace{-1 cm}

From a point of view in fields,

(A) \underline{Link Variables}  (Gauge Fields) are in  
either a blue or a red cell.

\vspace{0.2 cm}
(B) \underline{Plaquettes} are put only in 
a blue or red cell but \underline{not blank unit}.  

\vspace{0.2 cm}
(C) \underline{Site Variables} (Fermi Fields) are on 
 a boundary site of a blue or red cell. 
So fermi fields have dual property.  

Now we have to check not only the in-Cell condition but also inter-Cell
conditions.  There are many constraints but we can find a solution.
\section{SUMMARY AND DISCUSSIONS}
Let us summarize our work and give some discussions.  We presented a new
model for lattice super Yang-Mills theory.  After finding the SUSY 
invariant one-cell model, it is extended to the whole space.  Note that
the representation of a link variable, $U$, must be real as the SUSY
transformation of a fermi field.

What is the indication of our symmetry?  For example, Ward-Takahashi
identities can be written down.  A nontrivial example relates the field
strength and two-body fermion correlation.

We leave the investigation of the SUSY algebra including the actual
 spinor structure as a future problem.

Finally, the embedding pattern of the one-cell model allows the modulo 2
 translational invariance.  The pattern may imply an embedding of spinor
 structure to lattice.  In our forthcoming paper \cite{IKSSU}, some
 discussions will be given on this point.

\end{document}